# Initial Performance of CANGAROO-II 7m telescope

Hidetoshi Kubo[1], S.A. Dazeley[2], P.G. Edwards[3], S. Gunji[4], S. Hara[1], T. Hara[5], J. Jinbo[6], A. Kawachi[7], T. Kifune[7], J. Kushida[1], Y. Matsubara[8], Y. Mizumoto[9], M. Mori[7], M. Moriya[1], H. Muraishi[10], Y. Muraki[8], T. Naito[5], K. Nishijima[6], J.R. Patterson[2], M.D. Roberts[7], G.P. Rowell[7], T. Sako[11], K. Sakurazawa[1], Y. Sato[7], R. Susukita[12], T. Tamura[13], T. Tanimori[1], S. Yanagita[10], T. Yoshida[10], T. Yoshikoshi[7], and A. Yuki[8]

[1]*Department of Physics, Tokyo Institute of Technology, Meguro, Tokyo 152-8551, Japan*

[2]*Department of Physics and Mathematical Physics, University of Adelaide, South Australia 5005, Australia*

[3]*Institute of Space and Astronautical Science, Sagamihara, Kanagawa 229-8510, Japan*

[4]*Department of Physics, Yamagata University, Yamagata 990-8560, Japan*

[5]*Faculty of Management Information, Yamanashi Gakuin University, Kofu, Yamanashi 400-8575, Japan*

[6]*Department of Physics, Tokai University, Hiratsuka, Kanagawa 259-1292, Japan*

[7]*Institute for Cosmic Ray Research, University of Tokyo, Tanashi, Tokyo 188-8502, Japan*

[8]*STE Laboratory, Nagoya University, Nagoya, Aichi 464-860, Japan*

[9]*National Astronomical Observatory, Tokyo 181-8588, Japan*

[10]*Faculty of Science, Ibaraki University, Mito, Ibaraki 310-8521, Japan*

[11]*LPNHE, Ecole Polytechnique. Palaiseau CEDEX 91128, France*

[12]*Computational Science Laboratory, Institute of Physical and Chemical Research, Wako, Saitama 351-0198, Japan*

[13]*Faculty of Engineering, Kanagawa University, Yokohama, Kanagawa 221-8686, Japan*

**Abstract.** CANGAROO group constructed an imaging air Cherenkov telescope (CANGAROO-II) in March 1999 at Woomera, South Australia to observe celestial gamma-rays in hundreds GeV region. It has a 7m parabolic mirror consisting of 60 small plastic spherical mirrors, and the prime focus is equipped with a multi-pixel camera of 512 PMTs covering the field of view of 3 degrees. We report initial performance of the telescope.

## INTRODUCTION

CANGAROO group planned to construct an array of four 10m telescopes for stereoscopic observations of air Cherenkov lights [2]. In 1995 the construction of one imaging telescope was approved (CANGAROO-II project), however, due to the limit of the fund we decided to construct the telescope with a 7m mirror, of which frame and base can sustain a 10m mirror because we plan to expand the mirror from 7m to 10m in early 2000. In December 1998 all components of the telescope and the imaging camera were shipped to Australia. The construction of CANGAROO-II at Woomera, South Australia started in January 1999, and finished in the middle of March(Fig.1) [4]. In May the tuning of both electronics and trigger condition were carried out, and we started normal observations from June. Here we present the electronics system and the brief report of the performance as an imaging Cherenkov telescope. Performance of mirrors is described in detail by Kawachi et al. [1].

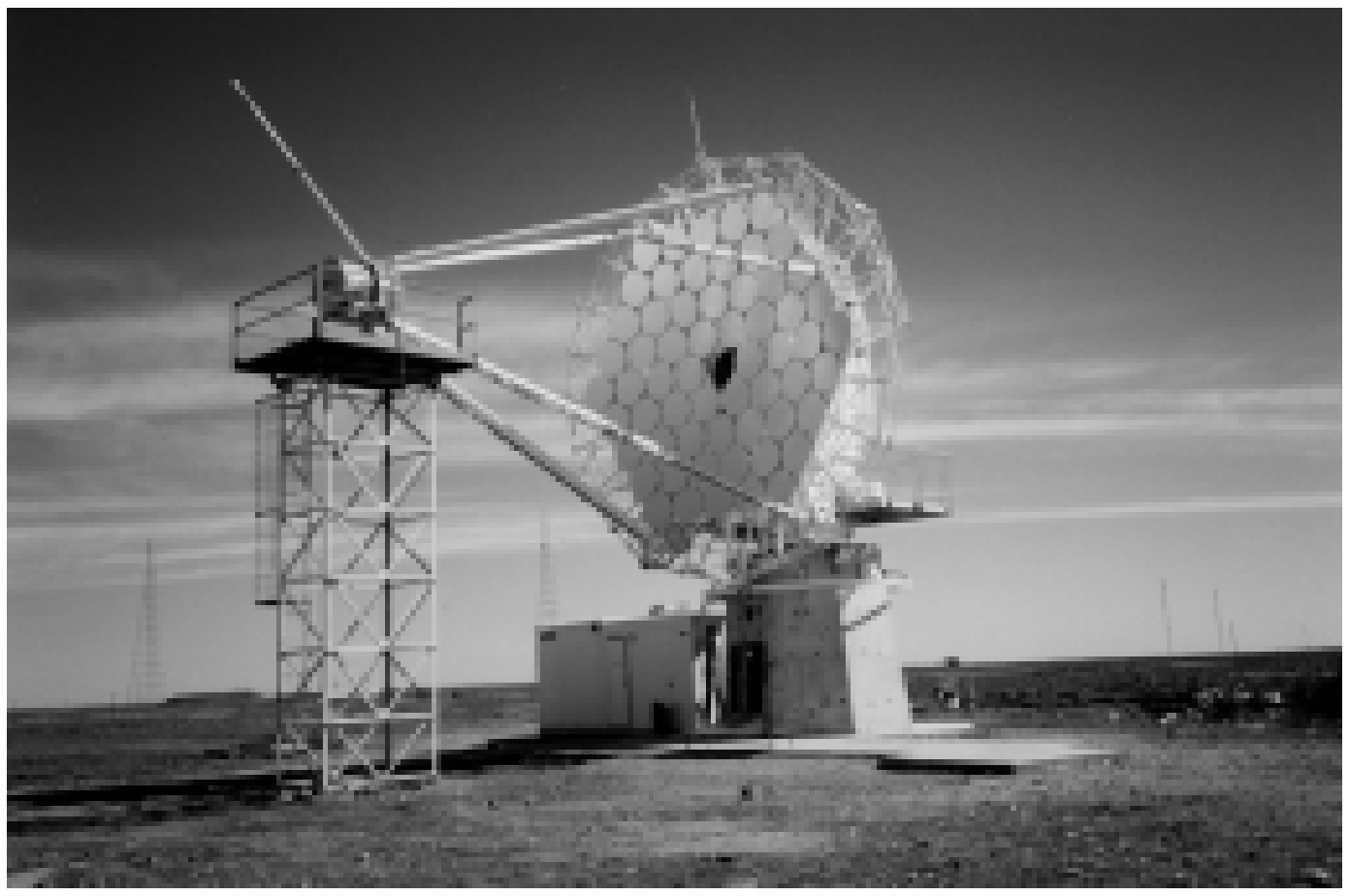

**FIGURE 1.** CANGAROO-II 7m telescope at Woomera, South Australia

## PERFORMANCE OF TELESCOPE

Figure 2 shows the front view of the camera attached in the focal plane, where that of 3.8m telescope is also shown for comparison. This camera consists of 512 pixels to cover a field of view (FOV) of diameter $\sim 3°$. Each pixel covers $0°.115 \times 0°.115$ (16mm $\times$ 16mm), and 13 mm $\phi$ photomultiplier (PMT: Hamamatsu R4124UV) was used as a pixel detector. The photocathode of this PMT has an area of 10 mm$\phi$ and covers about 35% of the FOV. The array of light collector were attached in front of the PMTs in order to increase the collection area of the camera by twice. Sixteen PMTs are housed in one module unit with a common bleeder circuit. PMTs are operated with a low gain of $\sim 10^5$ to avoid the gain drop for more than a few ten minutes after the passage of bright stars. Buffer amplifiers (LeCroy TRA402) are installed in the module unit to obtain the total gain of $\sim 10^7$ after amplification. The whole camera consists of these 32 module units.

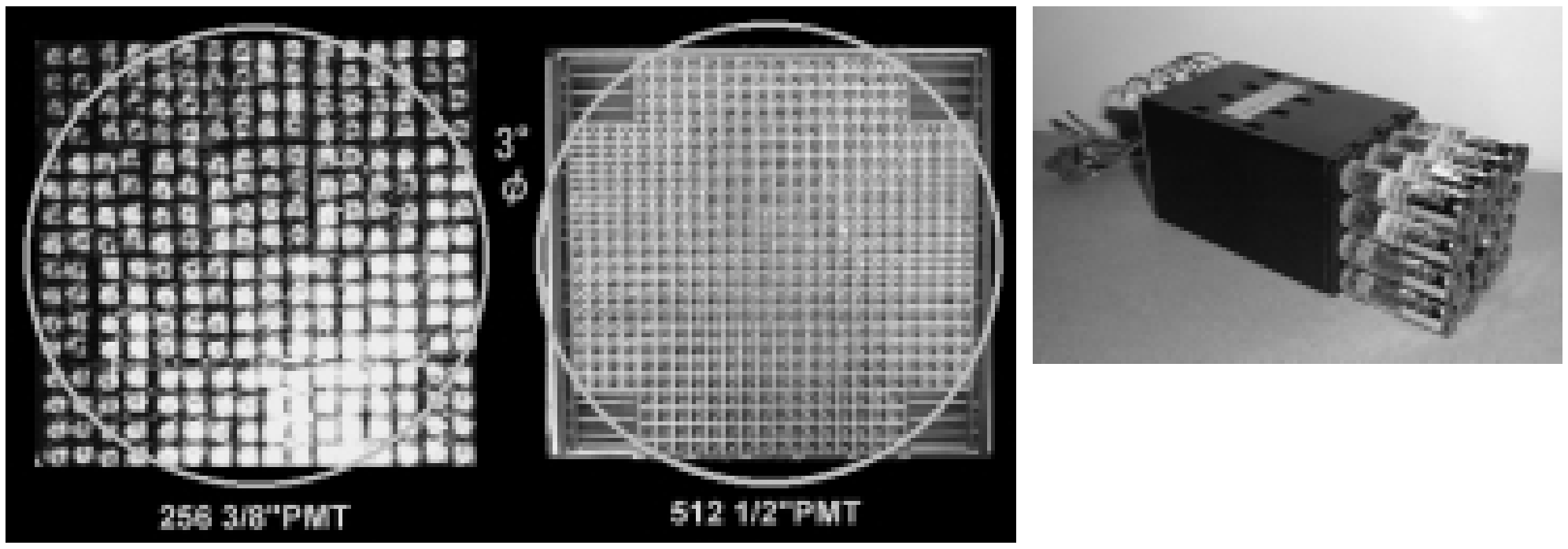

**FIGURE 2.** Front view of the cameras of the 3.8m telescope (left) and the new 7m telescope (middle). PMT module unit with 16 PMTs (right)

The signal from each PMT is transmitted through twisted cables of 36m in length to electronics circuits in the hut beside the base of the telescope. The arrival timing of lights detected with each PMT is recorded by multihit TDCs, and pulse heights are measured by charge ADCs in unit of a PMT base. The detail is described by Mori et al. [3]. At present triggers are generated when there are more than 4 PMTs whose pulse heights are larger than 4 photoelectrons, and the linear sum of any PMT module unit exceeds the night sky background fluctuation. Inner 16 of 32 PMT module units are concerned with the trigger, which covers a $\sim 1^\circ.8$ diameter of the FOV. In this condition triggers were generated at $\sim$ 10 Hz. Images taken with CANGAROO-II are shown in Fig. 3. Propagation of a shower triggered by hadron, and a ring triggered by muon are clearly seen. In 3.8m telescope, muon events were rarely detected since its detectable energy was relatively high ($\sim$ 1.5 TeV). The timing distribution of these events is shown in left panel of Fig. 4. The arrival timing of the muon event concentrates within 15 ns, on the other hand, that of the hadron event is distributed in 30 ns. The arrival timing accumulated for all events is also shown in right panel of Fig. 4. In case that the correction for the time jitter is not applied, the timing is distributed in 50 ns. After correction, the timing almost concentrates within 30 ns. The image parameter of off-source observation is shown in Fig. 5. The comparison with simulation is ongoing.

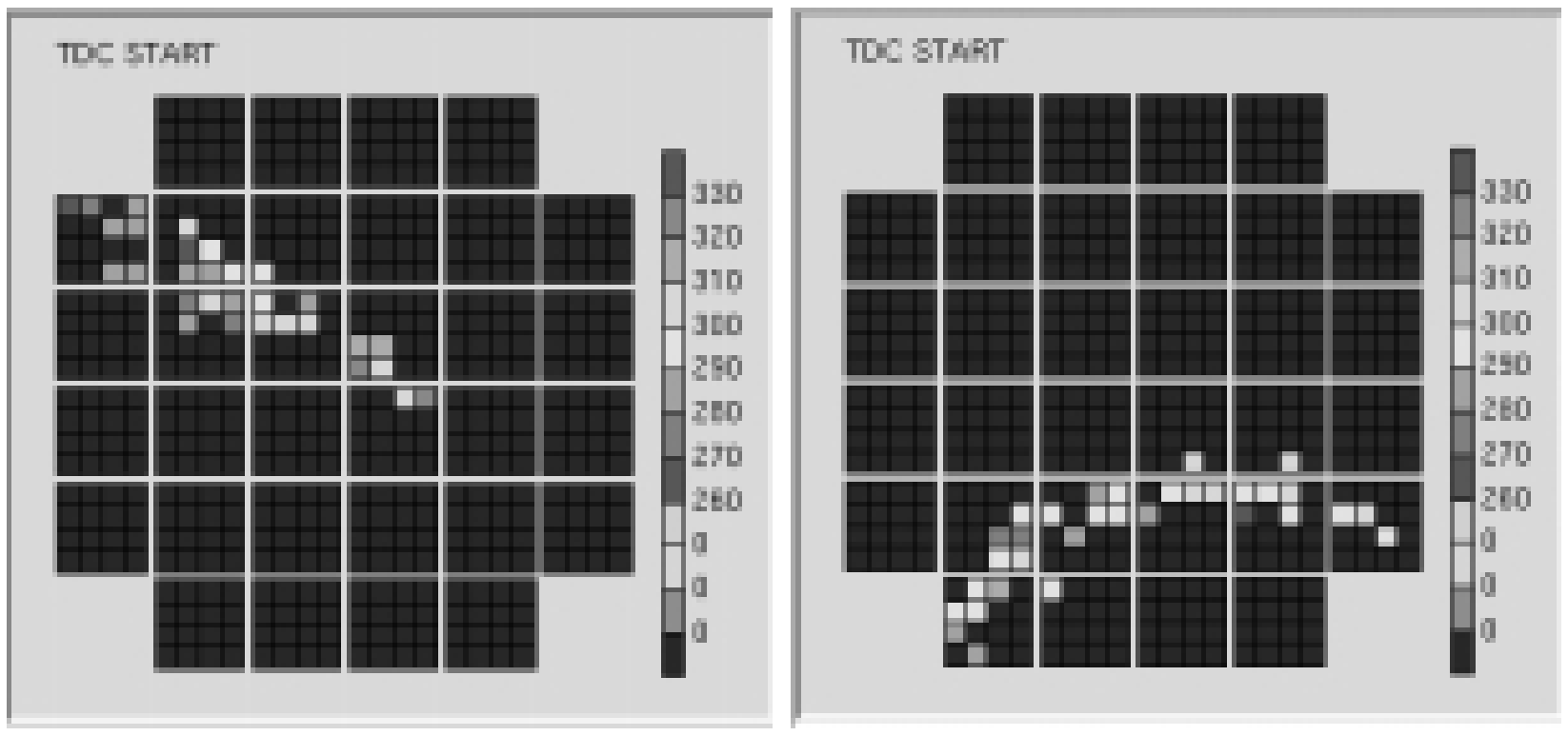

**FIGURE 3.** Images observed with CANGAROO-II, triggered by hadron(left) and muon(right).

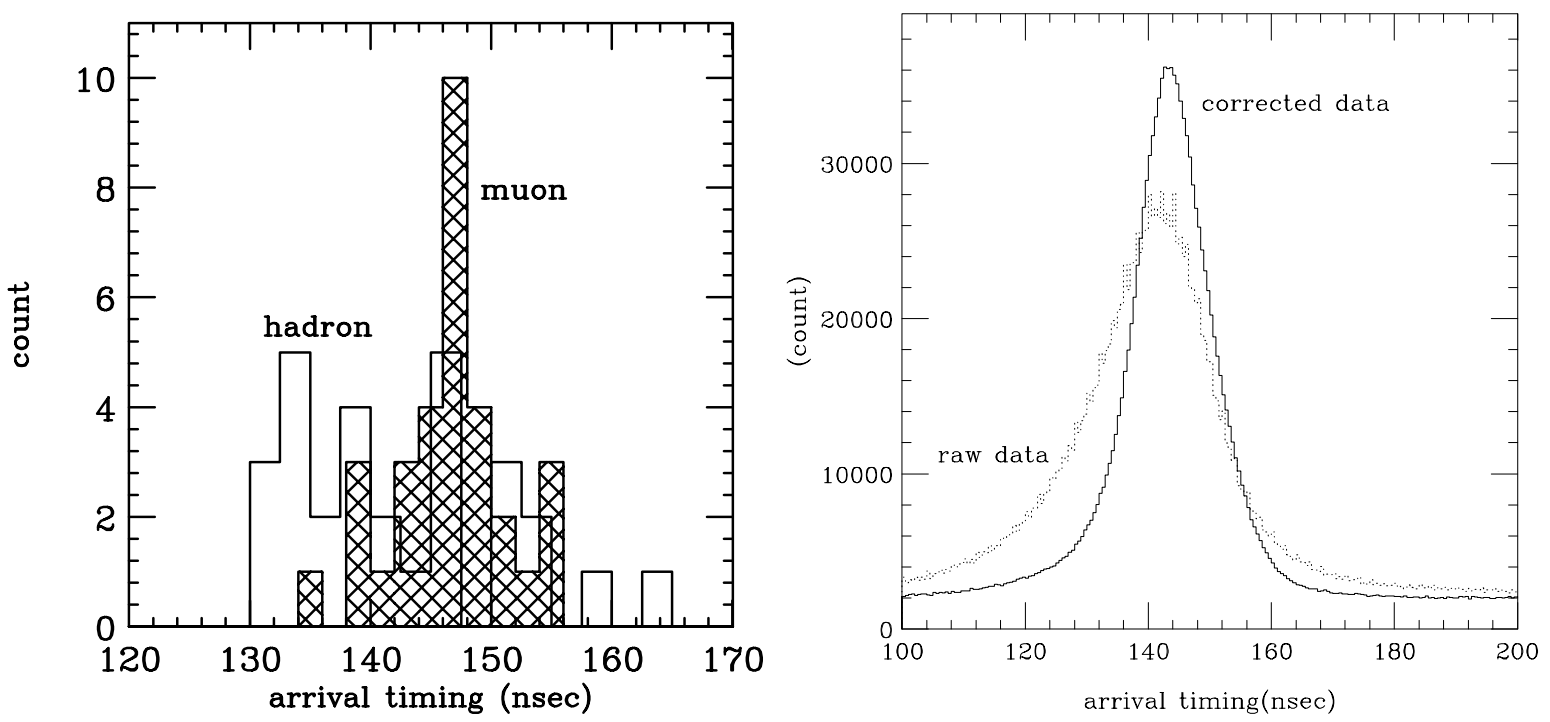

**FIGURE 4.** Arrival timing of hit PMTs for one event shown in Fig. 3(left), and for all events(right).

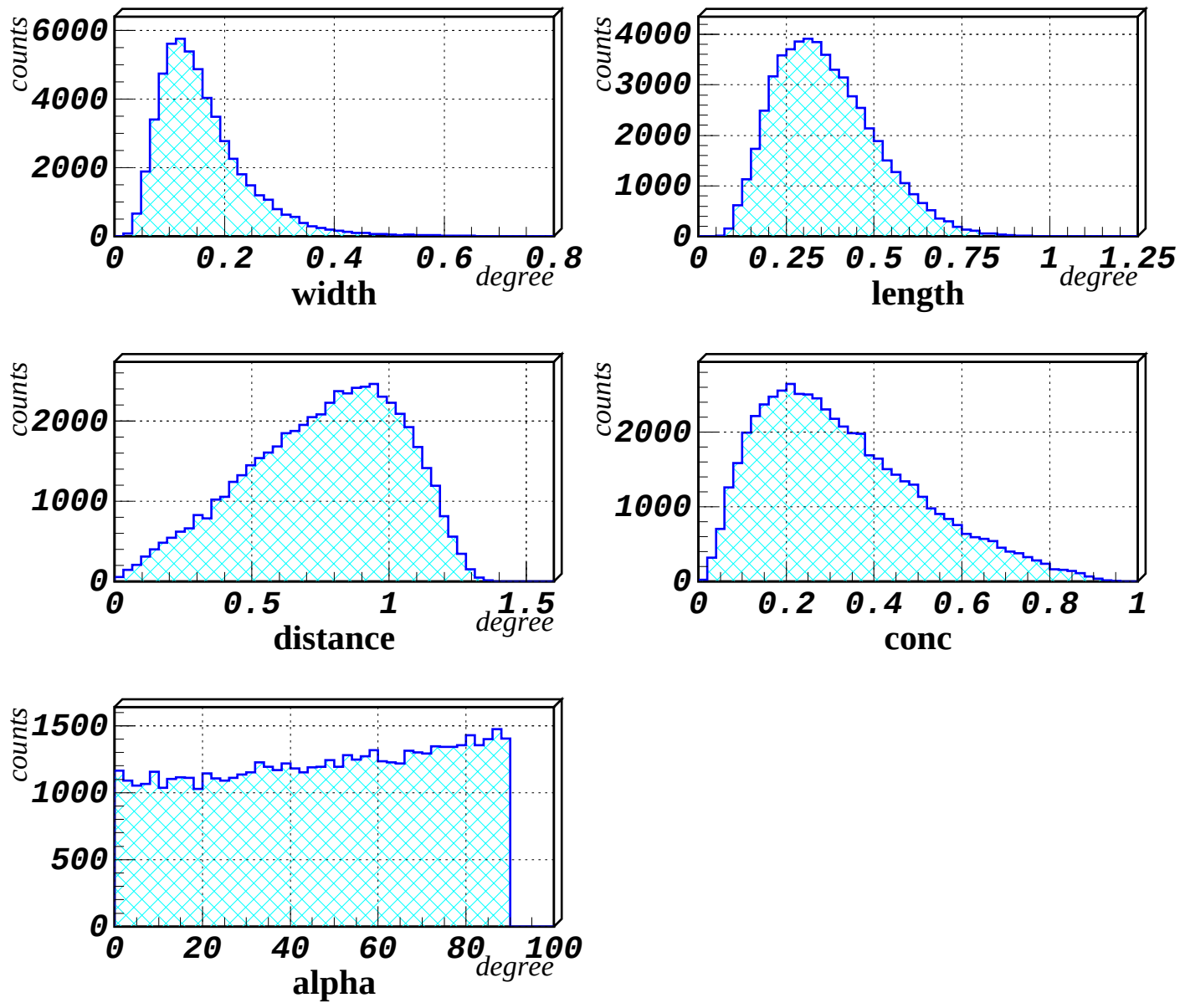

**FIGURE 5.** Image parameter of an off-source observation for 5 hours.

Figure 6 shows the distribution of the linear sum of all PMTs for triggered events. The slope becomes flatter as energy decreases. This might be due to the decrease of the detection efficiency at the trigger level for hadrons below ~1 TeV, which was expected for the large telescope having a fine imaging pixels from the simulation study as shown in Fig. 7. From comparison between the simulation and the observed spectrum, the energy threshold of CANGAROO-II 7m telescope is estimated to be ~300 GeV.

# SUMMARY

The construction of new CANGAROO-II 7m telescope has been completed in March 1999, and the telescope has detected many shower and muon events. The energy threshold is estimated to be ~ 300 GeV. The detailed study of the trigger condition and analysis of both on-source and off-source observations are ongoing.

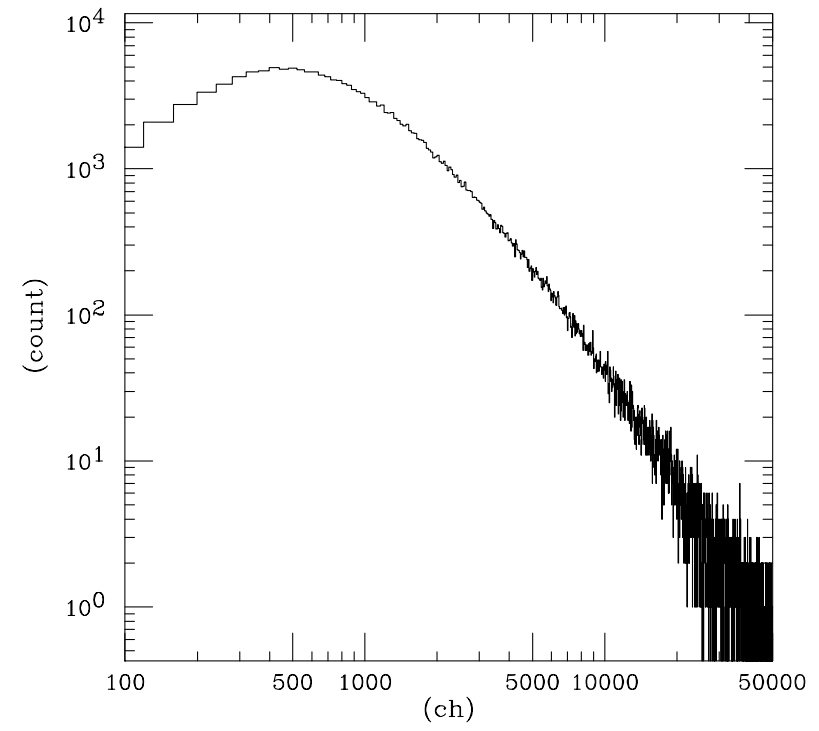

**FIGURE 6.** Observed distribution of the linear sum of all PMTs for triggered events

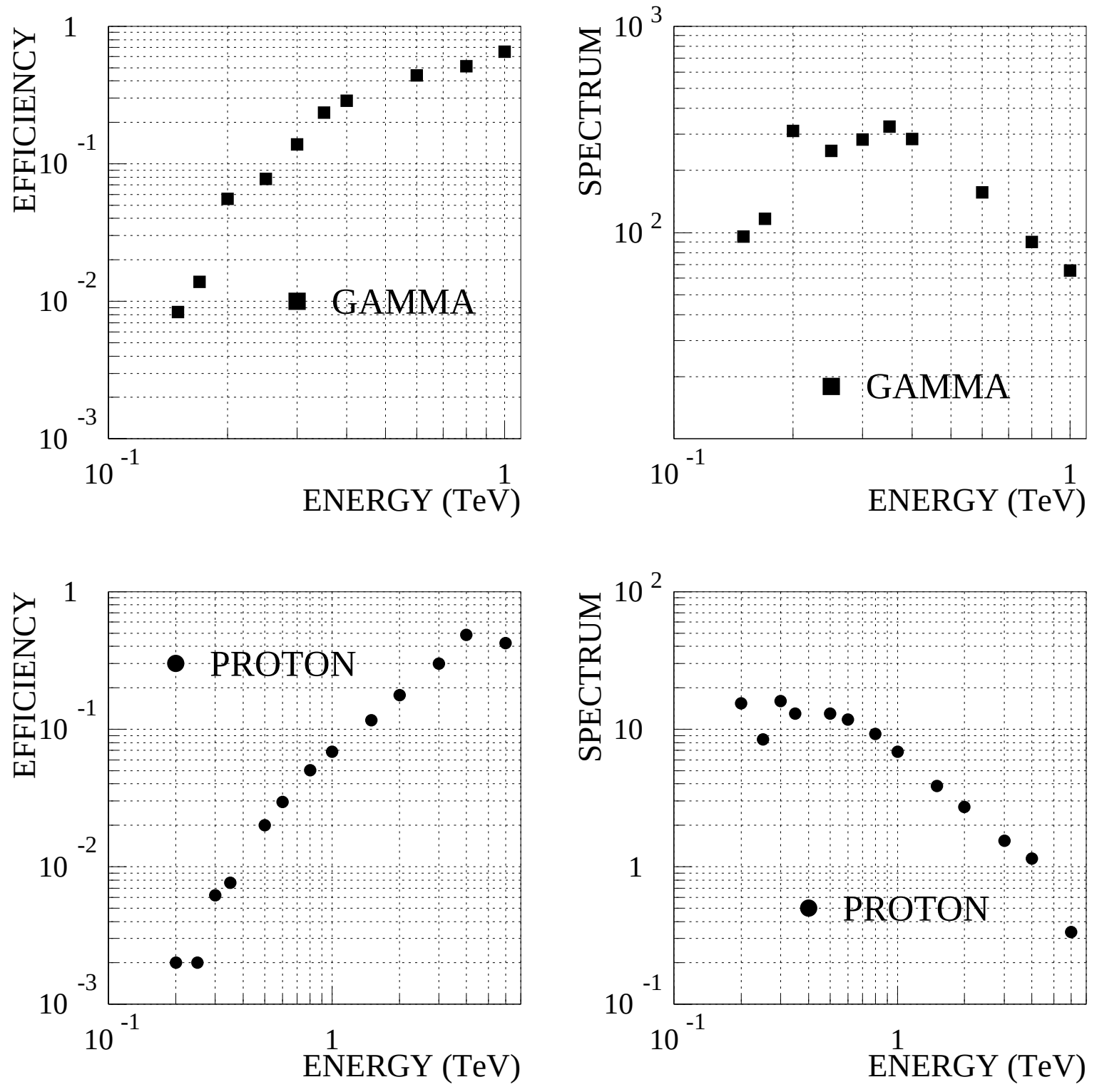

**FIGURE 7.** Simulated spectrum and efficiency for gamma-rays (upper) and protons (lower).